\pdfoutput=1
\documentclass[aps,prl, amsmath,amssymb,superscriptaddress,showkeys,reprint,preprintnumbers]{revtex4-1}
\usepackage{color}
\usepackage{aas_macros}
\usepackage{hyperref,breakurl}
\hypersetup{colorlinks=true, citecolor=blue, urlcolor=blue, linkcolor=blue}
\usepackage[usenames,dvipsnames]{xcolor}
\usepackage{dcolumn}
\usepackage{bm}
\usepackage{array}

\usepackage{amsmath,amssymb,latexsym,times}

\usepackage{todonotes}
 
\newcommand{\metacal}{\textsc{Metacalibration}}

\newcommand{\cosmosis}{\textsc{CosmoSIS}}

\newcommand{\redmagic}{\textsc{redMaGiC}}
\newcommand{\photoz}{photo-$z$}

\begin{document}

\title{Cosmological Constraints from Multiple Probes in the Dark Energy Survey}

\author{T.~M.~C.~Abbott}
\affiliation{Cerro Tololo Inter-American Observatory, National Optical Astronomy Observatory, Casilla 603, La Serena, Chile}
\author{A.~Alarcon}
\affiliation{Institut d'Estudis Espacials de Catalunya (IEEC), 08034 Barcelona, Spain}
\affiliation{Institute of Space Sciences (ICE, CSIC),  Campus UAB, Carrer de Can Magrans, s/n,  08193 Barcelona, Spain}
\author{S.~Allam}
\affiliation{Fermi National Accelerator Laboratory, P. O. Box 500, Batavia, IL 60510, USA}
\author{P.~Andersen}
\affiliation{School of Mathematics and Physics, University of Queensland,  Brisbane, QLD 4072, Australia}
\affiliation{University of Copenhagen, Dark Cosmology Centre, Juliane Maries Vej 30, 2100 Copenhagen O}
\author{F.~Andrade-Oliveira}
\affiliation{Instituto de F\'{i}sica Te\'orica, Universidade Estadual Paulista, S\~ao Paulo, Brazil}
\affiliation{Laborat\'orio Interinstitucional de e-Astronomia - LIneA, Rua Gal. Jos\'e Cristino 77, Rio de Janeiro, RJ - 20921-400, Brazil}
\author{J.~Annis}
\affiliation{Fermi National Accelerator Laboratory, P. O. Box 500, Batavia, IL 60510, USA}
\author{J.~Asorey}
\affiliation{Korea Astronomy and Space Science Institute, Yuseong-gu, Daejeon, 305-348, Korea}
\author{S.~Avila}
\affiliation{Institute of Cosmology and Gravitation, University of Portsmouth, Portsmouth, PO1 3FX, UK}
\author{D.~Bacon}
\affiliation{Institute of Cosmology and Gravitation, University of Portsmouth, Portsmouth, PO1 3FX, UK}
\author{N.~Banik}
\affiliation{Fermi National Accelerator Laboratory, P. O. Box 500, Batavia, IL 60510, USA}
% ...
\collaboration{DES Collaboration}

\noaffiliation

\date{\today}

\label{firstpage}

\begin{abstract}
The combination of multiple observational probes has long been advocated as a powerful technique to constrain cosmological parameters, in particular dark energy.
The Dark Energy Survey has measured 207 spectroscopically--confirmed Type Ia 
supernova lightcurves; the baryon acoustic oscillation feature; weak gravitational lensing; and galaxy clustering. 
Here we present combined results from these probes, deriving constraints on the equation of state, $w$, of dark energy and 
its energy density in the Universe.
Independently of other experiments, such as those that measure the cosmic microwave background, the probes from this single 
photometric survey rule out a Universe with no dark energy, finding $w=-0.80^{+0.09}_{-0.11}$. The geometry is shown to be consistent with a spatially flat Universe, and we obtain a constraint on the 
baryon density of $\Omega_b=0.069^{+0.009}_{-0.012}$ that is independent of early Universe measurements. 
These results demonstrate the potential power of large multi-probe photometric surveys and pave 
the way for order of magnitude advances in our constraints on properties of dark energy and cosmology over the next decade.
\end{abstract}


\preprint{DES-2018-0401}
\preprint{FERMILAB-PUB-18-585-AE}

\maketitle

\section{Introduction}\label{sec:intro}

The discovery of the accelerating Universe \cite{riess98,perlmutter99} revolutionized
 20th century cosmology by 
indicating the presence of a qualitatively new component in the
Universe that dominates the expansion in the last several billion years.  The nature of
dark energy --- the component that causes the accelerated expansion --- is unknown, and
understanding its properties and origin is one of the principal challenges in
modern physics.  Current measurements are consistent with
an interpretation of dark energy as a cosmological constant in General Relativity. 
Any deviation from this interpretation in space or time would constitute a 
landmark discovery in fundamental physics \cite{DETF}.

Dark energy leaves imprints on cosmological
observations, typically split into two regimes --- 1) it modifies the geometry of the Universe, increasing distances and volumes 
in the Universe over time via the accelerated expansion, and 2)
it suppresses the growth of cosmic structure. However, these effects can
 be mimicked by the variation of other cosmological parameters, including the dark matter density 
 and curvature, or other physical models and systematics that are degenerate within a single probe. Consequently, measuring dark energy properties 
requires a combination of cosmological probes that are sensitive to both classes of effects to break these parameter and model
degeneracies \cite{Frieman:2008sn,weinberg13,Huterer:2017buf}.

Historically, the most powerful cosmic probe has been
the cosmic microwave background
(CMB) \cite{Hu:2001bc,2013ApJS..208...19H,2018arXiv180706209P}, relic radiation from
the surface of last scattering only 400,000 years after the Big
Bang.  Low-redshift probes
measure the Universe over the last several billion years, when dark energy dominates the expansion.
Comparing or combining constraints between the CMB and lower redshift measurements 
requires us to extrapolate predictions to the present-day Universe starting from initial conditions over 13 billion years ago.
This is a powerful test of our models, but it requires precise, independent constraints from low-redshift experiments.
Low-redshift probes include Type Ia supernova (SNe Ia) measurements, which
treat the SNe Ia as standardizable candles and employ redshift and flux
measurements to probe the redshift-luminosity distance
relation \cite{2018ApJ...859..101S}; baryon acoustic oscillations (BAO), which
use a `standard ruler' scale in the cosmic density field, imprinted by sound waves at recombination, to probe several redshift-distance
combinations \cite{2014MNRAS.441...24A,2018ApJ...863..110B}; galaxy clustering, which measures the
density field up to some bias between galaxy density and the underlying dark
matter density, and redshift-space distortions (RSD) in the clustering \cite{Alam:2016hwk}; the counts of galaxy clusters,
representing the most extreme density peaks in the
Universe \cite{2011ARA&A..49..409A}; strong gravitational lensing \cite{2017MNRAS.468.2590S}; and weak gravitational lensing, 
which probes changes in the gravitational potential along the line of sight using coherent 
distortions in observed properties of galaxies or the CMB, e.g. to measure the dark and baryonic matter distribution \cite{2017arXiv171003235M}.

We report here the first results from the Dark Energy Survey (DES) combining precision probes of both geometry
and growth of structure that include BAO, SNe Ia, and weak lensing and galaxy clustering from a single experiment. 
DES has previously shown separate cosmological constraints using
weak lensing and galaxy clustering \cite{keypaper}, BAO \cite{DESY1BAO}, and SNe Ia \cite{snepaper}. We now combine these probes and begin to fully realize the
power of this multi-probe experiment to produce independent measurements of the properties of dark energy. 

The work presented here demonstrates our ability to extract and combine diverse cosmological 
observables from wide-field surveys of the evolved Universe.
Previous dark energy constraints 
have relied on combining the likelihoods of many separate and independent experiments to produce 
precise constraints on cosmological models including dark energy. 
For this traditional approach each experiment has performed an independent analysis 
to validate measurements and has separate calibration 
methodologies and requirements, thus ensuring that 
many potential systematics are uncorrelated between probes. 
The DES analysis presented here, however, uses a common set of 
both calibration methodologies and systematics modeling and marginalization across probes, 
which enables a consistently validated analysis. Perhaps most importantly, 
this common framework allows us to standardize requirements like blinding across these probes, which is essential to 
minimize the impact of experimenter bias \cite{2011arXiv1112.3108C}. This approach provides a very robust, precise cross-check of 
traditional multi-probe analyses, which currently provide tighter overall constraints.

The fundamental interest in understanding the nature of dark energy has spurred the development of 
multiple large photometric surveys that image the sky, capable of independently combining multiple cosmic probes. The current
generation of surveys includes the Hyper-Suprime Cam
survey (HSC) \cite{hsc}, the Kilo-Degree
Survey (KiDS) \cite{kids}, and the focus of this
work, DES \cite{des}.  
The next generation of these surveys will include the Large Synoptic
Survey Telescope (LSST) \cite{lsst}, a ground-based
telescope that will observe the entire southern hemisphere with very high
cadence, and space telescopes Euclid \cite{euclid} and the Wide-Field
InfraRed Survey Telescope (WFIRST) \cite{wfirst}. In parallel with imaging surveys, 
the distribution of galaxies measured by spectroscopic surveys (i.e., BOSS \cite{boss}, eBOSS \cite{eboss}, and the planned 4MOST \cite{4most}, DESI \cite{desi}, and PFS \cite{pfs} surveys) 
provides powerful constraints on the distance-redshift relation via BAO measurements and the growth of 
structure via redshift space distortions.
The union of these results over the following several years, 
and into the next decades, will ensure that we are able to take advantage of the benefits 
of multiple independent, self-consistent, and blinded multi-probe analyses like we present here for DES.

\section{Cosmic Probes}

\subsection{The Dark Energy Survey}
\label{sec:des}

DES cosmic probes span a wide range 
of redshifts up to $z\approx 1.3$, and include weak gravitational lensing and galaxy clustering due to large-scale structure \cite{keypaper}, 
SNe Ia \cite{snepaper}, and BAO \cite{DESY1BAO}. Each probe constrains dark energy independently and their 
combination is more powerful. These probes utilize a subset of data from DES 
taken during its first three observing seasons (Aug.~2013 to Feb.~2016). Spectroscopically confirmed SNe Ia are 
identified from images in all three seasons (DES Y3) in 27 deg$^2$ of repeated deep-field observations, while weak lensing and large-scale structure information is 
derived from images taken only in the first season (DES Y1), ending Feb. 2014 and covering 1321 deg$^2$ of the southern sky in 
$grizY$ filters. DES uses the 570-megapixel Dark 
Energy Camera (DECam \cite{decam}) at the Cerro Tololo Inter-American Observatory (CTIO) 4m Blanco 
telescope in Chile.  By the end of DES observations in January 2019, we 
anticipate an order of magnitude increase in the number of useable SNe, while the area of sky used for the other probes 
will increase by a factor of three to 5000 deg$^2$. Analysis of the later years of survey data is ongoing.

Data is processed through the DES Data Management system 
\cite{2012ApJ...757...83D,2011arXiv1109.6741S,2008SPIE.7016E..0LM,2018PASP..130g4501M}. 
This system detrends and calibrates the raw images, creates coadded images from individual exposures, 
and detects and catalogs astrophysical objects. This catalog is further cleaned and calibrated to create a 
high-quality (`Gold') object catalog \cite{y1gold} from which weak lensing and large-scale structure 
measurements are made. The deep fields are also processed through a separate difference imaging 
pipeline to identify transients \cite{Goldstein2015,Kessler2015}. The photometric and astrometric calibrations \cite{y1gold}
are common to all cosmology probes discussed below.

\subsubsection{Weak Gravitational Lensing and Large-Scale Structure}
\label{sec:3x2pt}

For weak gravitational lensing measurements, we use the measured shapes and positions of 26 million 
galaxies in the redshift range $0.2<z<1.3$, split into four redshift bins. The galaxy shapes are 
measured via the \metacal\ method \cite{HuffMandelbaum2017,SheldonHuff2017} using $riz$-band 
exposures \cite{shearcat}. 
Photometric redshifts for the objects are determined from a modified 
version of the BPZ method \cite{Benitez2000}, described and calibrated in Ref.~\cite{photoz}. 

For measurements of the angular galaxy clustering, we utilize the positions of a 
sample of luminous red galaxies that have precise photometric redshifts selected with the \redmagic\ 
algorithm \cite{2016MNRAS.461.1431R}. This results in a sample of 650,000 galaxies over the redshift 
range $0.15<z<0.9$, split into five narrow redshift bins. Residual correlations of number density with survey conditions 
in the \redmagic\ sample are calibrated in Ref.~\cite{wthetapaper}. 
The precise redshifts of \redmagic\ galaxies allow us to infer information about the 
more poorly constrained \photoz\ bias uncertainty in the weak lensing catalog. The \photoz\ calibration methodology is consistent between the weak lensing and \redmagic\ samples \cite{photoz,redmagicpz,xcorrtechnique,xcorr}.

We use measurements from each of these galaxy samples 
to construct a set of three two-point correlation function observables we label `3$\times$2pt'. These include the galaxy shear auto-correlation 
(cosmic shear), the galaxy position-shear cross-correlation (galaxy-galaxy lensing), and the galaxy position auto-correlation 
(galaxy clustering). The analysis was described in a series of papers that include the 
covariance and analysis framework \cite{2017MNRAS.470.2100K,methodpaper}, 
the measurements and validation \cite{simspaper,wthetapaper,gglpaper,shearcorr}, and the cosmological results \cite{keypaper}. 
We utilize the `3$\times$2pt' likelihood pipeline from this set of papers as implemented in \cosmosis\ \cite{cosmosis}. This 
combination of probes produces a tight constraint on the amplitude of matter clustering in the Universe
and on the properties of dark energy over the last six billion years.

\subsubsection{Type Ia Supernovae}\label{sec:sne}

The DES-SN sample is comprised of 207 spectroscopically confirmed SNe Ia 
in the redshift range $0.07<z<0.85$. The sample-building and analysis pipelines
are discussed in a series of papers that detail the SN Ia search and
discovery \cite{Goldstein2015,Kessler2015,2018PASP..130g4501M}; spectroscopic
follow-up \cite{DAndrea18}; photometry \cite{Brout18-SMP};
calibration \cite{Burke18,Lasker18}; simulations \cite{Kessler18}; and
technique of accounting for selection bias \cite{2017ApJ...836...56K,SK16}.  The analysis methodology and systematic
uncertainties are presented in Ref.~\cite{Brout18-SYS}. These results are used to constrain cosmology \cite{snepaper} and the Hubble constant \cite{Macaulay18}. 
In Refs.~\cite{snepaper,Brout18-SYS,Macaulay18} the DES-SN sample is combined with a `Low-$z$' ($z<0.1$) sample, which 
includes SNe from the Harvard-Smithsonian Center for
Astrophysics surveys \cite{Hicken09b,Hicken12} and the Carnegie Supernova
Project \cite{Contreras10}. Selection effects and calibration of these low-redshift samples
is discussed in \cite{2018ApJ...859..101S}. Here we fit for DES-SN alone, and 
only include the Low-$z$ sample for comparison to Ref.~\cite{snepaper}. 
We compute the SNe likelihood using the SNe module \cite{2018ApJ...859..101S} implemented in \cosmosis, which is able to reproduce the results in \cite{snepaper}.

\subsubsection{Baryon Acoustic Oscillations}\label{sec:bao}

A sample of 1.3 million galaxies from the DES Y1 `Gold' catalog in the redshift range $0.6 < z < 1.0$ was used to measure the BAO scale in the distribution of galaxies. Details of the galaxy sample selection are in Ref.~\cite{Crocce18}. Calibrations of the galaxy selection function are consistently derived for the BAO and `3$\times$2pt' samples.
This BAO measurement was presented in Ref.~\cite{DESY1BAO} and provides a likelihood for the ratio between the angular 
diameter distance to redshift 0.81, $D_A(z=0.81)$, and the sound horizon at the drag epoch, $r_d$. 
This analysis used 1800 simulations \cite{2018MNRAS.479...94A} and three methods to compute the galaxy clustering \cite{2017MNRAS.472.4456R,2018MNRAS.480.3031C,2018arXiv180710163C}.
The BAO likelihood is implemented in \cosmosis . The galaxy samples used in the `3$\times$2pt' angular clustering measurements and in the BAO 
analysis share a common footprint in the sky and overlap significantly in volume over the redshift range $0.6 < z < 0.9$, which will produce some non-zero correlation between the two measurements. However, the intersection of the galaxy populations is only about 14\% of the total BAO galaxy sample and we detect no significant BAO constraint when using the `3x2pt' galaxy clustering measurements. We thus ignore this negligible correlation when combining the two probes. 

\subsection{External Data for Comparison}\label{sec:ext}

We use external constraints that combine state-of-the-art CMB, SNe Ia, and spectroscopic BAO measurements to compare our results against. For the CMB data, 
we utilize full-sky temperature ($T$) and polarization ($E$- and $B$-mode) measurements from the 
Planck survey, combining $TT$ ($\ell \in [2,2508]$), and $EE$, $BB$ and $TE$ ($\ell \in [2,29]$)  
(commonly referred to as `TT+LowP') \cite{planck2015cosmo} with weak lensing measurements derived 
from the temperature data \cite{2016A&A...594A..15P}. We use the Planck 
likelihood from Ref.~\cite{2016A&A...594A..11P}.

For external SNe Ia measurements, we use the Pantheon compilation \cite{2018ApJ...859..101S}. 
Pantheon combines SNe Ia samples from Pan-STARRS1, SDSS, SNLS, various low-$z$ data sets, and HST. 
The Pantheon data set is based on the Pan-STARRS1 
Supercal algorithm \cite{2015ApJ...815..117S} that establishes a global calibration for the 13 different SNe Ia 
samples, with a total of 1048 SNe in $0.01<z<2.26$. 

Finally, external spectroscopic BAO measurements are taken from BOSS DR12 \cite{Alam:2016hwk}, the 
6dF Galaxy Survey \cite{Beutler:2011hx}, and the SDSS Main Galaxy Sample \cite{Ross:2014qpa}. These 
measurements of the BAO scale span a redshift range of $0.1<z<0.6$.

\section{Constraints on Dark Energy}\label{sec:de}

We present here a dark energy analysis that combines for the first time the DES probes described above. 
DES is able to strongly constrain dark energy models without the CMB by probing over a wide redshift 
range ($z\lesssim 1$) the growth of structure and distance-redshift relation, which are both sensitive to the presence of dark energy.
The dark energy equation of state $w$ relates
the pressure ($P$) to the energy density ($\rho$) of the dark energy fluid:
$w=P/\rho$, where $w=-1$ is equivalent to a cosmological constant $\Lambda$ in
the field equations. We probe the nature of dark energy in two ways: 1) we constrain the dark
energy density relative to the critical density today, $\Omega_{\Lambda}$, assuming that
dark energy takes the form of a cosmological constant and allowing non-zero
curvature (the $o$CDM model), and 2) we measure $w$ as a free parameter (the $w$CDM
model) with fixed curvature ($\Omega_k=0$). The total energy density of the Universe today is composed of the sum
of fractional components $1=\Omega_k+\Omega_m+\Omega_\Lambda$, where the 
components are: curvature ($\Omega_k$), the total matter ($\Omega_m$), and dark energy ($\Omega_\Lambda$). The radiation density is
assumed to be negligible over the redshift ranges probed by DES. 

In both $o$CDM and $w$CDM models, we explore the ability of DES to
constrain these properties of dark energy and compare this to the
state-of-the-art constraints combining measurements from many external
surveys. We follow the analysis methods and model definitions from Ref.~\cite{keypaper}, 
which includes varying the neutrino mass density in all models. External data are re-analyzed 
to make direct comparisons meaningful, including matching parameter choices and priors to the DES analysis. The cosmological parameters and their priors are slightly changed from  Ref.~\cite{keypaper} and listed in Table \ref{table:values}. Non-cosmological parameters and their priors are identical to Table 1 of Ref. \cite{keypaper}, with the absolute magnitude $-19.5<\mathcal{M}<-18.9$ for SNe. Cosmological parameters and the intrinsic alignment model (for `3$\times$ 2pt') are shared between probes. The joint posterior is the product of the individual posteriors of the three probes, which are assumed to be sufficiently independent at this precision, as motivated in the previous section.

\def\arraystretch{1.4}
\setlength{\tabcolsep}{4pt}
\begin{table}
\caption{Cosmological parameter constraints in the $o$CDM and $w$CDM
models using only DES data. We report the 1D peak of the posterior and asymmetric 68\% confidence limits. The marginalized parameters with informative priors (and prior ranges) are: the primordial perturbation amplitude $10^9 A_s\in [0.5,10.0]$, the Hubble constant $H_0\in [55, 90]$ \,{\rm km\, s$^{-1}$Mpc$^{-1}$}, the spectral 
index $n_s\in [0.87,1.07]$, and the neutrino mass density $\Omega_{\nu} h^2 \in [0.0006, 0.01]$.}
\label{table:values}
\begin{center}
\begin{tabular}{lcccc }
\hline
\hline
Parameter & $o$CDM & $w$CDM & $w$CDM (Ext) & Flat Prior \\ 
\hline
$\Omega_m$                                                  & $0.299^{+0.024}_{-0.020}$  &  $0.300^{+0.023}_{-0.021}$  & $0.303^{+0.007}_{-0.009}$ & [0.1, 0.9] \\
$\Omega_b$                                                   & $0.069^{+0.009}_{-0.012}$   & $0.064^{+0.013}_{-0.009}$  & $0.048^{+0.001}_{-0.001}$ & [0.03, 0.12] \\
$\Omega_k$                                                   & $0.252^{+0.095}_{-0.14}$     & 0 & 0 & [-0.1, 0.5] \\
$\Omega_\Lambda$                                       & $0.47^{+0.14}_{-0.12}$        & $0.700^{+0.021}_{-0.023}$  & $0.697^{+0.009}_{-0.007}$ & Derived \\
$w$                                                                 & $-1$                                      & $-0.80^{+0.09}_{-0.11}$       & $-1.02^{+0.03}_{-0.04}$ & [-2, -0.33] \\
$S_8$  							    & $0.801^{+0.028}_{-0.026}$  & $0.786^{+0.029}_{-0.019}$   & $0.814^{+0.016}_{-0.011}$ & Derived \\ 
\hline
\hline
\end{tabular}
\end{center}
\end{table}

\begin{figure}
\begin{center}
\includegraphics[width=\columnwidth]{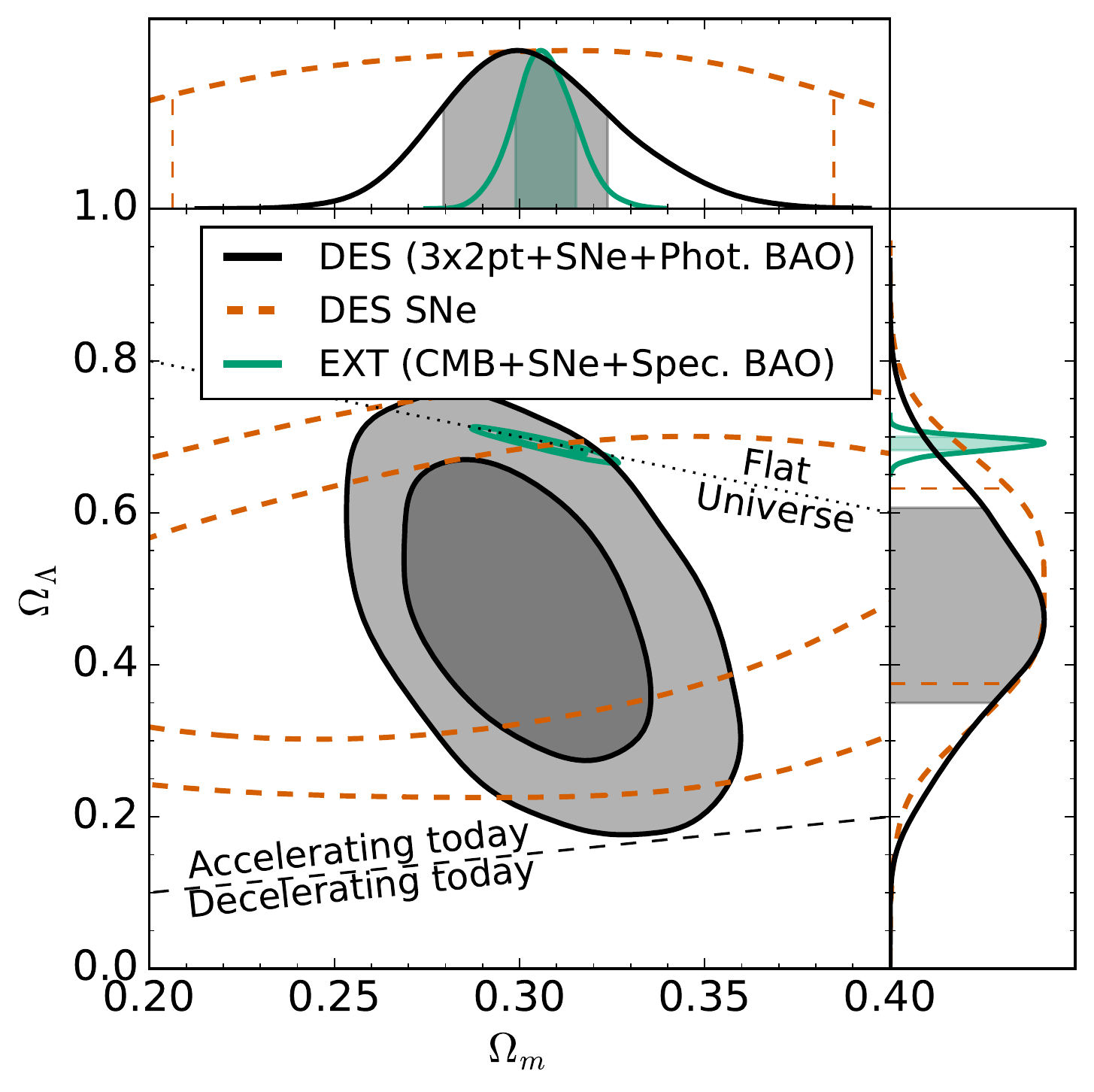}
\end{center}
\caption[]{
Constraints on the present-day dark energy density $\Omega_{\Lambda}$ and matter 
density $\Omega_m$, relative to the critical density, in an $o$CDM model with marginalized curvature and neutrino mass density. We compare the constraint from 
DES data alone (black contours), including information from weak gravitational lensing, large-scale structure, 
SNe Ia, and photometric BAO, to the best available external data (green contours), 
combining information from the CMB, SNe Ia, and spectroscopic BAO. We identify the flat model ($\Omega_k=0$) with a dotted line and distinguish accelerating and 
decelerating universes with a dashed line. Contours represent the 68\% and 95\% confidence limits (CL).
\label{fig:$o$CDM}}
\end{figure}

Figure~\ref{fig:$o$CDM} shows our constraints on $\Omega_\Lambda$ in the 
$o$CDM model, where $w=-1$. We combine our `3$\times$2pt', SNe Ia (without 
the external Low-$z$ sample), and photometric BAO measurements to constrain $\Omega_{\Lambda}$ 
and $\Omega_m$. This is compared to the constraint from the external data sets. The DES best-fit $\chi^2$ is 576 with 498 degrees of freedom (dof) \footnote{For further discussion of the $\chi^2$ value, see \cite{sndr} (Sec. 4 on Light Curve fitting and bias corrections).}.
Using DES data we are able to independently confirm the existence of a dark energy component 
in the Universe ($\Omega_\Lambda>0$) at $\sim$4$\sigma$ significance. This is the first time a photometric survey has independently made a significant
constraint on the energy density of both dark energy and dark matter without assuming a flat model based on early Universe constraints. It represents an important milestone
for future analyses from DES and surveys like Euclid, LSST, and WFIRST.

\begin{figure}
\begin{center}
\includegraphics[width=\columnwidth]{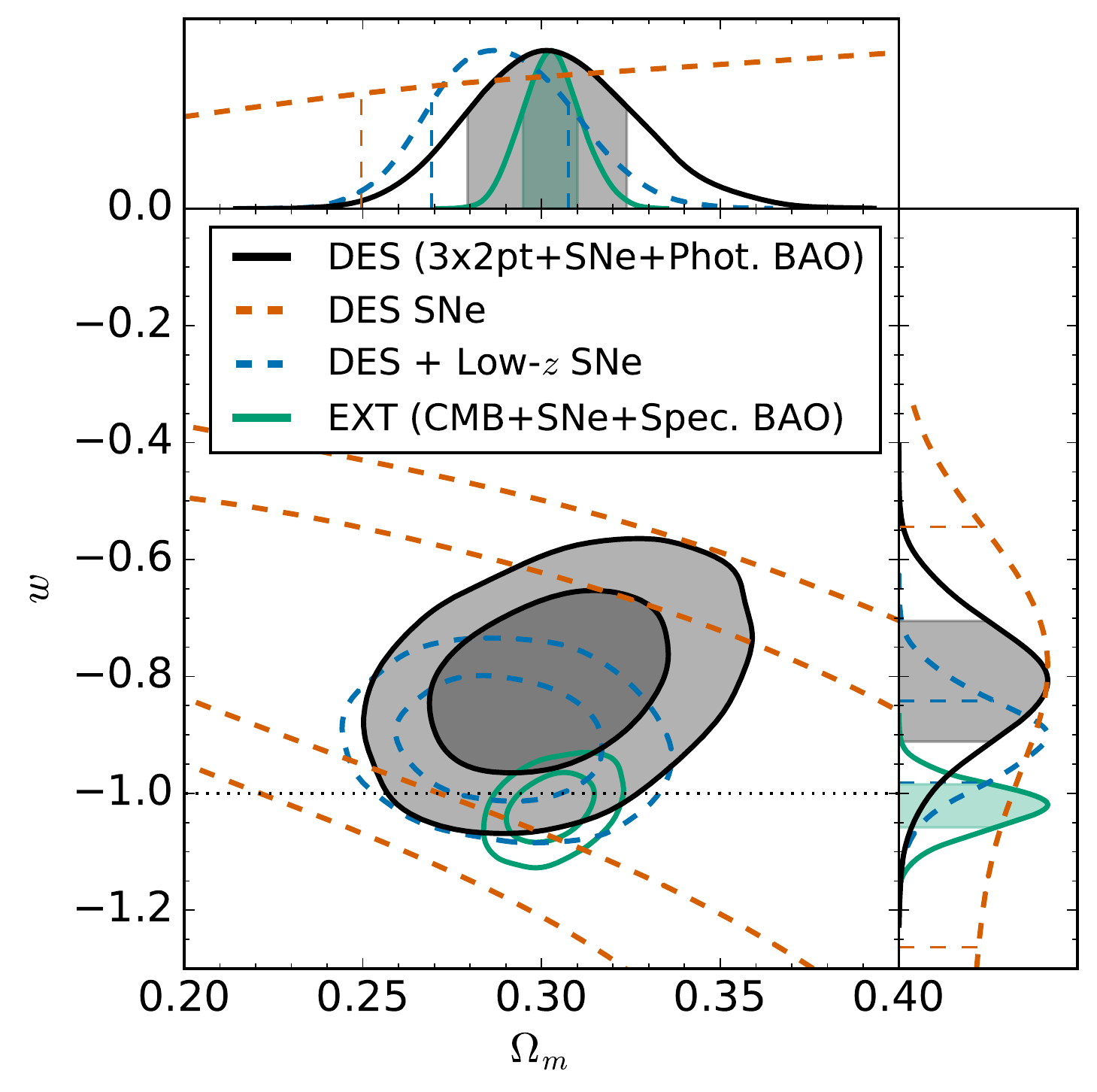}
\end{center}
\caption[]{Constraints on the dark energy equation of state $w$ and $\Omega_m$ in a 
$w$CDM model with fixed curvature ($\Omega_k=0$) and marginalized neutrino mass density. We compare constraints from the 
DES data alone (black contours) to the best available external data (green contours), as in Fig. ~\ref{fig:$o$CDM}, 
but also show the impact of including a low-redshift SNe Ia data set (Low-$z$) to 
anchor the DES SNe Ia as done in Ref.~\cite{snepaper} (blue contours). 
Each component of the DES analysis was fully blinded.
\label{fig:wcdm}}
\end{figure}

In Fig.~\ref{fig:wcdm}, we show the constraint on $w$ and $\Omega_m$, assuming the $w$CDM model. 
We show the same comparison with external
data as in Fig.~\ref{fig:$o$CDM}, but also include a case where we supplement
DES-discovered SNe Ia with the Low-$z$ SNe sample to anchor the SNe
redshift-distance relation at low redshift
following Ref.~\cite{snepaper}.  This low-redshift SNe anchor contributes significantly to both the DES+Low-$z$ and external constraints on $w$. 
In all cases, the existing data are consistent with a cosmological constant ($w=-1$). The DES best-fit $\chi^2$ is 577 with 498 dof.
This subset of the final DES data constrains $w$ to within a factor of three of the combined external constraint. This result illustrates the 
prospects for multiple independent, precise low-redshift constraints on dark energy from upcoming large-scale photometric experiments.

The constraints on all cosmological model parameters are summarized in Table~\ref{table:values}. 
Nuisance parameter constraints are not qualitatively changed from individual probe fits.
The DES-only `3$\times$2pt' and SNe data are consistent and individually contribute similar constraining power for $w$ and $\Omega_{\Lambda}$. 
In the $o$CDM model, DES constrains the total matter density to 7\% (68\% CL), 
the baryon density to 15\%, and the correlation amplitude to 3\%, described by $S_8\equiv \sigma_8 \sqrt{\Omega_m/0.3}$, 
where $\sigma_8$ measures the current-day clustering amplitude. The constraints are comparable in $w$CDM. Fixing $\Omega_k=0$, 
we find the $S_8$ constraint is improved by a factor of 1.2,  but there is otherwise no significant improvement in other parameters.
The parameter constraints beyond dark energy are driven by the `3$\times$2pt' measurement. In particular, 
the baryon density constraint is due to sensitivity to the shape of the 
matter power spectrum from baryon damping \footnote{This constraint on $\Omega_b$ is not prior driven, despite being degenerate with $A_s$ and $n_s$ (which is unconstrained).}.
The constraint on $\Omega_b$ from the CMB, by contrast, is also sensitive to the impact of baryons on the acoustic oscillations. 
Thus future low-redshift survey data will provide another avenue to test the predictions of our models from early Universe observations 
like the CMB with measurements of $\Omega_b$ from surveys like DES. 

\section{Outlook}\label{sec:conclusion}

The most precise constraints on dark energy properties require combining cosmological probes
that include information from both geometry and growth across cosmic history.
Thus far such diverse information was collected from different experiments,
which were subject to different calibration and systematic errors. We have combined for the first time in DES the 
purely cosmographic SN and BAO measurements with the growth-sensitive weak lensing and galaxy clustering 
measurements to independently place strong constraints on the nature of dark energy. These results share 
a common set of calibration frameworks and blinding policy across probes.
DES has independently constrained $\Omega_m$, $\Omega_b$, 
$\Omega_\Lambda$, $\sigma_8$, and $w$, while marginalizing over a free neutrino mass. 
We expect future DES results to provide a further factor of 2-4 improvement in these constraints due to increased area, 
depth, and number of SNe in the final analyses, which will then be followed by subsequent order of magnitude
 advances from more sensitive photometric surveys of the 2020s.

\section*{Acknowledgements}

Funding for the DES Projects has been provided by the DOE and NSF (USA), MEC/MICINN/MINECO (Spain), STFC (UK), HEFCE (UK), NCSA (UIUC), KICP (U. Chicago), CCAPP (Ohio State), MIFPA (Texas A\&M), CNPQ, FAPERJ, FINEP (Brazil), DFG (Germany) and the Collaborating Institutions in the Dark Energy Survey.

The Collaborating Institutions are Argonne Lab, UC Santa Cruz, University of Cambridge, CIEMAT-Madrid, University of Chicago, University College London, 
DES-Brazil Consortium, University of Edinburgh, ETH Z{\"u}rich, Fermilab, University of Illinois, ICE (IEEC-CSIC), IFAE Barcelona, Lawrence Berkeley Lab, 
LMU M{\"u}nchen and the associated Excellence Cluster Universe, University of Michigan, NOAO, University of Nottingham, Ohio State University, University of 
Pennsylvania, University of Portsmouth, SLAC National Lab, Stanford University, University of Sussex, Texas A\&M University, and the OzDES Membership Consortium.

Based in part on observations at Cerro Tololo Inter-American Observatory, National Optical Astronomy Observatory, which is operated by the Association of 
Universities for Research in Astronomy (AURA) under a cooperative agreement with the National Science Foundation.

The DES Data Management System is supported by the NSF under Grant Numbers AST-1138766 and AST-1536171. 
The DES participants from Spanish institutions are partially supported by MINECO under grants AYA2015-71825, ESP2015-66861, FPA2015-68048, SEV-2016-0588, SEV-2016-0597, and MDM-2015-0509, 
some of which include ERDF funds from the European Union. IFAE is partially funded by the CERCA program of the Generalitat de Catalunya.
Research leading to these results has received funding from the European Research
Council under the European Union's Seventh Framework Program (FP7/2007-2013) including ERC grant agreements 240672, 291329, and 306478.
We  acknowledge support from the Australian Research Council Centre of Excellence for All-sky Astrophysics (CAASTRO), through project number CE110001020, and the Brazilian Instituto Nacional de Ci\^encia
e Tecnologia (INCT) e-Universe (CNPq grant 465376/2014-2).

This manuscript has been authored by Fermi Research Alliance, LLC under Contract No. DE-AC02-07CH11359 with the U.S. Department of Energy, Office of Science, Office of High Energy Physics. The United States Government retains and the publisher, by accepting the article for publication, acknowledges that the United States Government retains a non-exclusive, paid-up, irrevocable, world-wide license to publish or reproduce the published form of this manuscript, or allow others to do so, for United States Government purposes.

This research used resources of the National Energy Research Scientific Computing Center, a DOE Office of Science User Facility supported by the Office of Science of the U.S. Department of Energy under Contract No. DE-AC02-05CH11231. This work also used resources on the CCAPP condo of the Ruby Cluster at the Ohio Supercomputing Center \cite{OhioSupercomputerCenter1987}. Plots in this manuscript were produced partly with \textsc{Matplotlib} \cite{Hunter:2007}, and it has been prepared using NASA's Astrophysics Data System Bibliographic Services.

\bibliographystyle{apsrev4-1}
\bibliography{short}

\label{lastpage}

\end{document}